\documentclass[9pt,technote]{IEEEtran}

\usepackage{graphicx}
\usepackage{color}
\usepackage{amsmath,amsfonts,amssymb,mathrsfs}

\begin{document}

\newtheorem{theorem}{Theorem}
\newtheorem{lemma}{Lemma}
\newtheorem{remark}{Remark}
\newtheorem{definition}{Definition}
\newtheorem{proposition}{Proposition}

\title{Convergence of Distributed Randomized
PageRank Algorithms}

\author{Wenxiao Zhao, Han-Fu Chen, and Hai-Tao Fang
\thanks{The authors are with Key Laboratory of Systems and Control, Academy of Mathematics and Systems Science,
Chinese Academy of Sciences and National Center for Mathematics and
Interdisciplinary Sciences, Chinese Academy of Sciences, China.}
\thanks{
Email addresses: {\sf wxzhao@amss.ac.cn} (Wenxiao Zhao), {\sf
hfchen@iss.ac.cn} (Han-Fu Chen), {\sf htfang@iss.ac.cn} (Hai-Tao
Fang).}
\thanks{The research of Han-Fu Chen is supported by NSF of
China (NSFC) under Grants No. 61273193, 61120106011, and 61134013.
The research of Hai-Tao Fang is supported by NSFC under Grant No.
61174143. The research of Wenxiao Zhao is supported by NSFC under
Grants No. 61104052, 61273193, and 61134013.}}

\date{}
\maketitle

\begin{abstract}
The PageRank algorithm employed by Google quantifies the importance
of each page by the link structure of the web. To reduce the
computational burden the distributed randomized PageRank algorithms
(DRPA) recently appeared in literature suggest pages to update their
ranking values by locally communicating with the linked pages. The
main objective of the note is to show that the estimates generated
by DRPA converge to the true PageRank value almost surely under the
assumption that the randomization is realized in an independent and
identically distributed (iid) way. This is achieved with the help of
the stochastic approximation (SA) and its convergence results.
\end{abstract}


\section{Introduction}

The PageRank algorithm employed by Google quantifies the importance
of each page by the link structure of the web and it has achieved a
great success as a commercial searching engine. Let us first recall
the PageRank problem presented in \cite{BrinPage}\cite{Tempo2010}.
Consider a web with $n$ pages. The web is modeled by a direct graph
$\mathscr{G}=(\mathscr{V},\mathscr{E})$, where
$\mathscr{V}=\{1,2,\cdots,n\}$ is the index set of the pages and
$\mathscr{E}\subset \mathscr{V}\times \mathscr{V}$ is the set of
links representing the structure of the web. If $(i,j)\in
\mathscr{E}$, then page $i$ has an outgoing link to page $j$.
Without losing generality, we assume $n>2$.

Denote by $S_j$ the set of those pages which have incoming links
from page $j,$ and by $n_j$ the number of pages in $S_j.$ Thus we
have associated with the graph $\mathscr{G}$ a link matrix
\begin{align}
A=[a_{ij}]_{n\times n},~~a_{ij}=
\begin{cases}
\frac{1}{n_j},~j\in \mathscr{L}_i,\\
0,~\mathrm{otherwise},
\end{cases}\label{1}
\end{align}
where $\mathscr{L}_i=\{j:(j,i)\in \mathscr{E}\}.$ It is clear that
$\sum^n_{i=1}a_{ij}$ equals either $1$ or $0.$

The importance of a page $i$ is characterized by its PageRank value
$x^*_i\in [0,1],~~i\in \mathscr{V}.$ Let us assume
$\sum_{i=1}^nx_i^*=1$. The basic idea of the PageRank algorithm is
that a page which has links from important pages is also important.
Mathematically, this suggests to define the PageRank value of page
$i$ by
\begin{align}
x^*_i=\sum\limits_{j\in \mathscr{L}_i}\frac{x^*_j}{n_j},\label{2}
\end{align}
 or equivalently, to define $x^*=[x_1^*,\cdots,x_n^*]^T$ from the following linear algebraic equation
\begin{align}
x^*=Ax^*,~~x_i^*\in[0,1].\label{3}
\end{align}
The normalization condition $\sum_{i=1}^nx_i^*=1$ is possible to be
met because if $x^*$ satisfies (\ref{3}) then $\lambda x^*$ with
$\lambda\in (0,1]$ also satisfies (\ref{3}).

However, since the huge growing size of Internet, 8 billion pages as
reported in \cite{Tempo2010}, the matrix $A$ is of dimension 8
billion$\times$8 billion, the computation of the PageRank value
$x^*$ becomes a problem. It is reported in \cite{Meyer} that the
classical centralized method, such as the Power Method
\cite{BrinPage}, is rather time-consuming. In this regards, several
other approaches, such as the adaptive computation method
\cite{Kamvar}, distributed randomized method
\cite{Tempo2009}\cite{Tempo2010}\cite{Tempo2011}\cite{Tempo2012}\cite{Zhu},
asynchronous iteration method \cite{Jager}\cite{Kollias}, etc., have
been proposed. See also \cite{Andersen}, \cite{Avrachenkov},
\cite{Broder}, and \cite{Juditsky} among others.

According to the DRPA approach introduced in
\cite{Tempo2009}\cite{Tempo2010}\cite{Tempo2011}\cite{Tempo2012},
the pages can update their PageRank values by locally communicating
with the linked pages and the computation load required by this
approach is rather mild. It has been shown that the estimates
generated by this kind of algorithms converge to the true PageRank
value in the mean-square sense. The main objective of the note is to
show that the estimates generated by DRPA converge to the true
PageRank value with probability one under the assumption that the
randomization is realized in an iid way. To achieve this, some
results from SA algorithm \cite{Chen2002} are applied.

The rest of the note is arranged as follows. In Section II, DRPA is
introduced and the main results of the note are presented. The
convergence analysis is given in Section III. The DRPA in a more
general case is discussed in Section IV and some concluding remarks
are addressed in Section V.

{\em Notations.} Denote by $(\Omega,\mathscr{F},\mathbb{P})$ the
basic probability space and by $\omega\in\Omega$ a sample in the
probability space. A probability vector $x=[x_1\cdots x_n]^T\in
\mathbb{R}^n$ is defined as $x_i\geq0,~i=1,\cdots,n$ and
$\sum_{i=1}^nx_i=1$, while a stochastic matrix $A=[a_{ij}]\in
\mathbb{R}^{n\times n}$ is defined as $a_{ij}\geq0,~i,j=1,\cdots,n$
and $\sum_{i=1}^na_{ij}=1,~j=1,\cdots,n$. Denote by $S\in
\mathbb{R}^{n\times n}$ and $\mathbf{1}\in \mathbb{R}^{n}$ the
matrix and vector with all entries being $1$. We say that a matrix
or a vector is positive if all the entries are positive. By $\|x\|$
we denote the Euclidean norm of a vector $x\in \mathbb{R}^n$.
Finally, by $b\bar\in B$ we mean that the element $b$ does not
belong to the set $B$.

\section{Distributed Randomized PageRank Algorithm}

We recall some basic results in PageRank computation. Notice that in
the real world there exist nodes, for example, the dangling nodes,
which have no outgoing links to other nodes and thus correspond to
zero columns of the link matrix $A$. To avoid the computational
difficulty caused by this, the following assumption A1) is often
made on the matrix $A$.
\begin{itemize}
\item[A1)] $A\in \mathbb{R}^{n\times n}$ is a stochastic matrix.
\end{itemize}

%In applications, this can be satisfied by slightly modifying the
%link structure of the web.

From (\ref{3}) it is clear that the PageRank value of the web is the
eigenvector corresponding to eigenvalue $1$ of the matrix $A$. In
order the eigenvalue $1$ to have multiplicity 1, the following
technique is adopted in \cite{BrinPage} and \cite{Tempo2010}. Define
the matrix $M\in \mathbb{R}^{n\times n}$ by
\begin{align}
M\triangleq(1-\alpha)A+\alpha\frac{S}{n},\label{4}
\end{align}
where $\alpha\in(0,1)$.

\begin{lemma}
(\cite{BrinPage}\cite{Tempo2010}) If A1) holds, then the following
assertions take place.
\begin{itemize}
\item[i)] $M$ is a positive stochastic matrix, whose eigenvalue $1$ is with multiplicity $1,$
and all eigenvalues of $M$ are in the closed unit disk;

\item[ii)] The eigenvectors $\overline{x}$ and $\overline{\overline{x}}$ of $M$ corresponding to eigenvalue $1$
satisfy $\overline{x}=-\overline{\overline{x}},$ and one of them is
positive.
\end{itemize}
\end{lemma}

\begin{definition} (\cite{BrinPage})
The PageRank value $x^*$ of web $\mathscr{G}$ is defined by
\begin{align}
x^*=Mx^*,~x^*_i\in[0,1],~\sum\limits_{i=1}^nx^*_i=1.\label{5}
\end{align}
\end{definition}

%Due to the huge size of the web, it is difficult to directly
%compute all the eigenvalues and eigenvectors of the matrix $M$.

A widely used solution of the PageRank problem (\ref{5}) is the
Power Method (\cite{Horn}) which is recursively computed:
\begin{align}
x_{k+1}=Mx_k=(1-\alpha)Ax_k+\frac{\alpha}{n}\mathbf{1}\label{6}
\end{align}
with $x_0\in \mathbb{R}^n$ being a probability vector.

\begin{lemma}(\cite{BrinPage}\cite{Tempo2010})
For the Power Method (\ref{6}) the following convergence takes
place:
\begin{align}
x_k\mathop{\longrightarrow}\limits_{k\to\infty}x^* \label{7}
\end{align}
for any probability vector $x_0$.
\end{lemma}

DRPA considered in \cite{Tempo2009}\cite{Tempo2010}\cite{Tempo2012}
makes the link matrices $\{A_i\}$, to be defined below, to be sparse
and thus greatly simplifies the computation.

Consider the web $\mathscr{G}=(\mathscr{V},\mathscr{E}).$  The basic
idea of DRPA is as follows: At time $k$, page $i$ updates its
PageRank value by locally communicating with the pages which have
incoming links from page $i$ and/or outgoing links to page $i,$ and
page $i$ which takes the above action is determined in a random
manner. To be precise, DRPA is given by
\begin{align}
x_{1,k+1}=(1-\alpha_1)A_{\theta(k)}x_{1,k}
+\frac{\alpha_1}{n}\mathbf{1}\label{8}
\end{align}
with $x_{1,0}$ being an arbitrary probability vector,
$\alpha_1=\frac{2\alpha}{n-\alpha(n-2)}$ and the link matrix
\begin{align}
(A_i)_{jl}\triangleq
\begin{cases}
a_{jl},~~\mathrm{if}~j=i~\mathrm{or}~l=i\\
1-a_{il},~~\mathrm{if}~j=l\neq i\\
0,~~\mathrm{otherwise}
\end{cases}\label{9}
\end{align}
for $i=1,\cdots,n,$ and where
 $\{\theta(k)\}_{k\geq0}$ is assumed to be a sequence of iid random variables
with probability
\begin{align}
\mathbb{P}\{\theta(k)=i\}=\frac{1}{n},~i=1,\cdots,n.\label{10}
\end{align}
It is clear that matrices $\{A_i\}$ are sparse.

\begin{lemma}(\cite{Tempo2010})
If A1) holds and $\alpha_1=\frac{2\alpha}{n-\alpha(n-2)}$, then
\begin{itemize}
\item[i)] the matrix $M_1$ defined by $M_1\triangleq(1-\alpha_1)EA_{\theta(k)}
+\frac{\alpha_1}{n}S$ is a positive stochastic matrix and satisfies
\begin{align}
M_1=\frac{\alpha_1}{\alpha}M+\left(1-\frac{\alpha_1}{\alpha}\right)I,
~\mbox{and}~Ex_{1,k+1}=M_1Ex_{1,k};\label{11}
\end{align}
\item[ii)]  the average $\overline{x}_{1,k+1}$ of
$\{x_{1,0},\cdots,x_{1,k}\}$
\begin{align}
\overline{x}_{1,k+1}=\frac{1}{k+1}\sum\limits_{l=0}^{k}x_{1,l}\label{12}
\end{align}
converges to $x^*,$ the PageRank value of the web
$(\mathscr{V},\mathscr{E})$, in the mean square sense
$E\|\overline{x}_{1,k}-x^*\|^2\mathop{\longrightarrow}\limits_{k\to\infty}0.$
\end{itemize}
\end{lemma}

The matrix $A_i$ describes the local link structure of page $i$. The
choice of $\alpha_1=\frac{2\alpha}{n-\alpha(n-2)}$ is to make $M_1$
to take the form
$M_1=\frac{\alpha_1}{\alpha}M+\left(1-\frac{\alpha_1}{\alpha}\right)I$
so that $M_1$ and $M$ share the common eigenvector corresponding to
their biggest eigenvalue. This enables $Ex_{1,k}$ generated from
$Ex_{1,k+1}=M_1Ex_{1,k}$ converges to $x^*$. However, $Ex_{1,k}$ is
unavailable and thus the average type algorithm (\ref{12}) is
adopted. In the following, we consider the almost sure convergence
of $\{\overline{x}_{1,k}\}_{k\geq0}$.

%In \cite{Tempo2010} it is shown that the sequence
%$\{x_{1,k}\}_{k\geq0}$ generated by (\ref{8}) is ergodic and hence
%it cannot converge to the equilibrium point of equation $x=Mx$. This
%is why the average-type algorithm (\ref{12}) is adopted.

Notice that algorithm (\ref{12}) can be written in a recursive way:
\begin{align}
\overline{x}_{1,k+1}=\overline{x}_{1,k}-
\frac{1}{k+1}(\overline{x}_{1,k}-x_{1,k}).\label{13}
\end{align}

Recall that the SA algorithm (or the Robbins-Monro algorithm
\cite{Chen2002}\cite{Kushner})
\begin{align}
z_{k+1}=z_k+\gamma_k(f(z_k)+\varepsilon_{k+1}),~k\geq0,\label{14}
\end{align}
is used to search the roots of $f(z)=0,$ where $\gamma_k$ is the
stepsize, $f(z)$ is the unknown function with the observation
$f(z_k)+\varepsilon_{k+1}$, and $\varepsilon_{k+1}$ is the
observation noise at time $k+1$. Comparing (\ref{13}) and
(\ref{14}), we find that (\ref{13}) is precisely an SA algorithm
with the unknown function
$f(\overline{x}_{1,k})=-(\overline{x}_{1,k}-x^*)$ valued at
$\overline{x}_{1,k}$ and the observation noise
$\varepsilon_{k+1}=-(x^*-x_{1,k})$.
%\begin{align}
%f(\overline{x}_{1,k})=-(\overline{x}_{1,k}-x^*),\label{15}
%\end{align}
%\begin{align}
%\varepsilon_{k+1}=-(x^*-x_{1,k}).\label{16}
%\end{align}
This observation motivates us to establish the almost sure
convergence of $\{\overline{x}_{1,k}\}_{k\geq0}$ by the convergence
analysis for SA \cite{Chen2002}. Indeed, we have the following
results to be proved in Section III.

\begin{theorem}\label{Thm1}
If A1) holds and $\alpha_1=\frac{2\alpha}{n-\alpha(n-2)}$, then the
estimate generated by DRPA (\ref{8}) and (\ref{13}) converges to the
true PageRank value almost surely:
\begin{align}
\overline{x}_{1,k}-x^*\mathop{\longrightarrow}\limits_{k\to\infty}0~~\mathrm{a.s.}\label{17}
\end{align}
\end{theorem}

\begin{remark}\label{rem1}
By the boundedness of $\{\overline{x}_{1,k}\}$, the strong
consistency of $\overline{x}_{1,k}$ implies its convergence in the
mean square sense.
\end{remark}

\begin{theorem}\label{Thm2}
For $\overline{x}_{1,k}$ generated from (\ref{13}) and with
$\alpha_1=\frac{2\alpha}{n-\alpha(n-2)}$, the following convergence
rate takes place for any $\epsilon\in \left(0,\frac{1}{2}\right)$:
\begin{align}
\|\overline{x}_{1,k}-x^*\|=o\left(\frac{1}{k^{\frac{1}{2}-\epsilon}}\right)~~
\mathrm{a.s.}\label{34}
\end{align}
\end{theorem}

\section{Convergence Analysis}

We first present the basic convergence results of stochastic
approximation algorithm with expanding truncations (SAAWET), which
the proof for Theorems 1 and 2 is essentially based on. More results
on SAAWET can be found in \cite{Chen2002}.

Assume the root $z^0$ of $g(\cdot):~\mathbb{R}^N\to \mathbb{R}^N$ is
a singleton, i.e., $g(z^0)=0$.

Let $\{M_k\}_{k\geq0}$ be a positive sequence increasingly diverging
to infinity and let $\{z_k\}_{k\geq0}$ be given by the following
algorithms:
\begin{align}
\nonumber
z_{k+1}=&\left[z_k+\gamma_ky_{k+1}\right]I_{[\|z_k+\gamma_ky_{k+1}\|\leq
M_{\sigma_k}]}\\
&+z^*I_{[\|z_k+\gamma_ky_{k+1}\|>
M_{\sigma_k}]},\label{a1}\\
\sigma_k=&\sum\limits_{i=1}^{k-1}I_{[\|z_i+\gamma_iy_{i+1}\|>
M_{\sigma_i}]},~~~~\sigma_0=0,\label{a2}\\
y_{k+1}=&g(z_k)+\varepsilon_{k+1}.\label{a3}
\end{align}

We need the following conditions.

\begin{itemize}
\item[ C1)] $\gamma_k>0,~\gamma_k\mathop{\longrightarrow}
\limits_{k\to\infty}0,~\mbox{and}~\sum_{k=1}^{\infty}\gamma_k=\infty.$

\item[C2)] There is a continuous differentiable function
$v(\cdot):\mathbb{R}^N\to \mathbb{R}$ such that $
\sup\limits_{\delta\leq \|z-z^0\|\leq \Delta}\triangledown
v(z)^{T}g(z)<0,~~~~\forall~\Delta>\delta>0. $ Further, $z^*$ used in
(\ref{a1}) is such that $
v(z^*)<\inf\limits_{\|z\|=c_0}v(z)~~\mbox{for~some}~c_0>0~~\mbox{and}~\|z^*\|<c_0.
$

\item[C3)] For the sample path $\omega$ under consideration,
$$
\lim\limits_{T\to 0}\limsup\limits_{k\to\infty}\frac{1}{T}
\left\|\sum\limits_{i=n_k}^{m(n_k,T_k)}\gamma_i\varepsilon_{i+1}\right\|
=0,~~~~\forall~T_k\in[0,T]
$$
for any $\{n_k\}$ such that $\{z_{n_k}\}$ converges, where
$m(k,T)=\max\left\{m:\sum\limits_{j=k}^m \gamma_j\leq T\right\}$.

\item[C4)] $g(\cdot)$ is measurable and locally bounded.
\end{itemize}

\begin{proposition}\label{prop1}
(Theorem 2.2.1 in \cite{Chen2002}) Assume that C1), C2), and C4)
hold. Then $z_k\mathop{\longrightarrow}\limits_{k\to\infty}z^0$ for
those $\omega$ for which C3) holds.
\end{proposition}

\begin{remark}\label{rem2}
(Remark 2.2.6 in \cite{Chen2002}) If we know that $\{z_k\}$ given by
an SAAWET algorithm evolves in a subspace of $\mathbb{R}^N$, then it
suffices to verify C2) in the subspace in order the corresponding
convergence of $\{z_k\}$ to hold.
\end{remark}

\begin{remark}\label{rem3}
Compared with the classical SA algorithm, such as the
Robbins-Monro's algorithm, the conditions required for convergence
of SAAWET are significantly weaker. For details we refer to Chapters
1 and 2 of \cite{Chen2002}.
\end{remark}

For the convergence rate of SAAWET, the following conditions are to
be used.

\begin{itemize}
\item[ C1')] $\gamma_k>0,~\gamma_k\mathop{\longrightarrow}
\limits_{k\to\infty}0,~\sum_{k=1}^{\infty}\gamma_k=\infty,$ and
$\frac{\gamma_k-\gamma_{k+1}}{\gamma_k\gamma_{k+1}}\mathop{\longrightarrow}
\limits_{k\to\infty}\sigma\geq0$.

\item[C3')] For the sample path $\omega$ under consideration, the
noise $\{\varepsilon_k\}$ in C3) can be decomposed into two parts
$\varepsilon_k=\varepsilon_k'+\varepsilon_k''$ such that
$$
\sum\limits_{k=1}^{\infty}\gamma_k^{1-\delta}\varepsilon'_{k+1}<\infty,~~
\varepsilon_{k+1}''=O(\gamma^{\delta}_k)
$$
for some $\delta\in(0,1]$.

\item[C4')] $g(\cdot)$ is measurable and locally bounded and is
differentiable at $z^0$ such that as $z\to z^0$
$$
g(z)=F(z-z^0)+o(\|z-z^0\|).
$$
The matrices $F$ and $F+\sigma\delta I$ are stable, where $\sigma$
and $\delta$ are given in C1') and C3'), respectively.
\end{itemize}

\begin{proposition}\label{prop2}
(Theorem 3.1.1 in \cite{Chen2002}) Assume C1'), C2), and C4') hold.
Then on those sample paths for which C3') holds, $z_k$ converges to
$z^0$ with the following convergence rate:
$$
\|z_k-z^0\|=o(\gamma_k^{\delta}),
$$
where $\delta$ is the one given in C3').
\end{proposition}

To apply Propositions 1 and 2, the key point is to verify the
convergence of random series like
$\sum_{k=1}^{\infty}\gamma_k\varepsilon_{k+1}$. For this the
following proposition plays an important role.

Define $\{\alpha_k=[k^a]\}_{k\geq0}$ for some $a>1$, where $[c]$
denotes the integer part of the number $c,$ and define
$I_0(0)\triangleq\{\alpha_0,\alpha_1,\alpha_2,\cdots\}$,
$I_j(i)\triangleq\{\alpha_j+i,\alpha_{j+1}+i,\cdots\}$ and
$I_j\triangleq\bigcup_{i=i_1(j)}^{i_2(j)}I_j(i)$, where
$i_1(j)\triangleq\alpha_j-\alpha_{j-1},~i_2(j)\triangleq\alpha_{j+1}-\alpha_{j}-1$.

\begin{proposition}\label{prop3}(\cite{HuChen2006})
\begin{itemize}
\item[i)] The sets
$\{I_j(i)\}_{j,i}$ defined above are  disjoint and
$I_0(0)\bigcup\Bigg\{\bigcup_{j=1}^{\infty}\Big[\bigcup_{i=i_1(j)}^{i_2(j)}I_j(i)\Big]\Bigg\}
=\{0,1,2,3,\cdots\}$.

\item[ii)] Let $\{\zeta_k\}$ be a sequence of random vectors with
zero mean and $\sup_{k}E\|\zeta_k\|^2<\infty$. If for any fixed $j$
and $i:~i_1(j)\leq i \leq i_2(j)$, the subsequence $\{\zeta_k:k\in
I_j(i)\}$ is composed of mutually independent random variables with
possible exception of a finite number of $\zeta_k$, then
\begin{align*}
\sum\limits_{k=1}^{\infty}\frac{1}{k^s}\zeta_k<\infty~~\mathrm{a.s.}
\end{align*}
for all $s>\frac{3}{2}-\frac{1}{a}$.
\end{itemize}
\end{proposition}

We now proceed to prove Theorem 1.

{\em Proof of Theorem 1:} As discussed in Section II, algorithm
(\ref{13}) can be rewritten as
\begin{align}
\overline{x}_{1,k+1}=\overline{x}_{1,k}+
\frac{1}{k+1}(-(\overline{x}_{1,k}-x^*)+e_{1,k+1}+e_{2,k+1}),\label{18}
\end{align}
where $e_{1,k+1}=-(x^*-Ex_{1,k})$ and
$e_{2,k+1}=-(Ex_{1,k}-x_{1,k}).$

By the fact that both $x_{1,k}$ and $\overline{x}_{1,k}$ are
probability vectors for all $k\geq0$, thus
$\|\overline{x}_{1,k}\|\leq 1$ and the SA algorithm (\ref{18}) is in
fact an SAAWET algorithm whose estimate sequence evolves in a
bounded subspace of $\mathbb{R}^n$. So by Proposition \ref{prop1}
and Remark \ref{rem2}, for (\ref{17}), we only need to find a
Lyapunov function to meet C2) and to verify the noise condition C3).

Define $f(x)=-(x-x^*)$ and the Lyapunov function
$V(x)\triangleq\|x-x^*\|^2$. It follows that
\begin{align}
\sup\limits_{\delta<\|x-x^*\|<\Delta,\|x\|\leq1}\triangledown
V(x)^Tf(x)<0,\label{19}
\end{align}
for any $0<\delta<\Delta$. Hence assumption C2) holds.

So, by Proposition 1, to prove (\ref{17}) it suffices to show that
\begin{align}
Ex_{1,k}-x^*\mathop{\longrightarrow}\limits_{k\to\infty}0\label{23}
\end{align}
and
\begin{align}
\sum\limits_{k=0}^{\infty}
\frac{1}{k+1}(x_{1,k+1}-Ex_{1,k+1})<\infty~~\mathrm{a.s.}\label{24}
\end{align}

By Lemma 3, $M_1$ and $M$ share the same eigenvector $x^*$
corresponding to eigenvalue $1$. Then by (\ref{11}) and Lemma 2, we
know that (\ref{23}) holds. In what follows we show that (\ref{24})
takes place.

Define the matrix
\begin{align}
\Phi(k,j)\triangleq
\begin{cases}
A_{\theta(k)}A_{\theta(k-1)}\cdots A_{\theta(j)},~~\mathrm{if}~j\leq
k,\\
I,~~\mathrm{if}~j=k+1.
\end{cases}\label{26}
\end{align}

Then equation (\ref{8}) can be rewritten as
\begin{align}
\nonumber x_{1,k+1}=&(1-\alpha_1)^{k+1}\Phi(k,0)x_{1,0}\\
&+\frac{\alpha_1}{n}\sum\limits_{l=1}^{k+1}(1-\alpha_1)^{k+1-l}\Phi(k,l)\mathbf{1},\label{27}
\end{align}
from which it follows that
\begin{align}
\nonumber &x_{1,k+1}-Ex_{1,k+1}\\
\nonumber=&
(1-\alpha_1)^{k+1}\Big(\Phi(k,0)x_{1,0}-E\Phi(k,0)x_{1,0}\Big)\\
&+ \frac{\alpha_1}{n}\sum\limits_{l=1}^{k+1}(1-\alpha_1)^{k+1-l}
\Big(\Phi(k,l)\mathbf{1}-E\Phi(k,l)\mathbf{1}\Big).\label{49}
\end{align}

By noticing that $x_{1,0}$ is a probability vector and
$\{A_{\theta(k)}\}$ are stochastic matrices, it is clear that
\begin{align}
\sum\limits_{k=1}^{\infty}\frac{1}{k+1}(1-\alpha_1)^{k+1}
\Big(\Phi(k,0)x_{1,0}-E\Phi(k,0)x_{1,0}\Big)<\infty~~\mathrm{a.s.}\label{28}
\end{align}
Thus, for (\ref{24}) it remains to show that
\begin{align}\label{29}
\sum\limits_{k=1}^{\infty}\frac{1}{k+1}\xi_{k+1}<\infty~~\mathrm{a.s.},
\end{align}
where
\begin{align*}
\xi_{k+1}\triangleq\sum\limits_{l=1}^{k+1}(1-\alpha_1)^{k+1-l}
(\Phi(k,l)\mathbf{1}-E\Phi(k,l)\mathbf{1}).
\end{align*}

For any fixed $a>1$, define $\alpha_k\triangleq [k^a],~k\geq 0.$
Further, as for Proposition 3 define
$I_0(0)\triangleq\{\alpha_0,\alpha_1,\alpha_2,\cdots\}$,
$I_j(i)\triangleq\{\alpha_j+i,\alpha_{j+1}+i,\cdots\}$ and
$I_j\triangleq\bigcup_{i=i_1(j)}^{i_2(j)}I_j(i)$, where
$i_1(j)\triangleq\alpha_j-\alpha_{j-1},~i_2(j)\triangleq\alpha_{j+1}-\alpha_{j}-1$.

By Proposition 3, the sets $\{I_j(i)\}_{j,i}$ are disjoint and
$I_0(0)\bigcup\Big\{\bigcup_{j=1}^{\infty}\Big[\bigcup_{i=i_1(j)}^{i_2(j)}I_j(i)\Big]\Big\}
=\{0,1,2,3,\cdots\}$.

Take $\tau\in\left(0,1-\frac{1}{a}\right)$ and define
\begin{align}
\overline{\xi}_{k+1}\triangleq\sum\limits_{l=k+1-[k^{\tau}]}^{k+1}(1-\alpha_1)^{k+1-l}
(\Phi(k,l)\mathbf{1}-E\Phi(k,l)\mathbf{1}).\label{30}
\end{align}

Notice that $\{\xi_k\}$ is not mutually independent. For any fixed
$j\geq 1$ and
$i\in[\alpha_j-\alpha_{j-1},\alpha_{j+1}-\alpha_{j}-1]$, let us
consider the set $\{\overline{\xi}_{k+1}:k+1\in I_j(i)\}$ and show
that $\{\overline{\xi}_{k+1},~k+1\in I_j(i)\}$ are mutually
independent with possible exception of a finite number of
$\overline{\xi}_{k+1}.$

If $k+1\in I_j(i)$, then
$\overline{\xi}_{k+1}=\overline{\xi}_{[m^a]+i}$ for some integer
$m.$ By definition $\overline{\xi}_{[m^a]+i}$ is measurable with
respect to the $\sigma$-algebra
$\sigma\{\theta([m^a]+i-1),\cdots,\theta([m^a]+i-[([m^a]+i-1)^{\tau}])\}.$

In the set $\{\overline{\xi}_{k+1}:k+1\in I_j(i)\},$ the random
vector $\overline{\xi}_{[(m-1)^a]+i}$ is with subscript neighboring
with $[(m-1)^a]+i.$ Since $\{\theta(k)\}$ is iid, for the mutual
independence of random vectors in $\{\overline{\xi}_{k+1}:k+1\in
I_j(i)\}$ it suffices to show that $\overline{\xi}_{[m^a]+i}$ and
$\overline{\xi}_{[(m-1)^a]+i}$ are independent. It is clear that for
this it suffices to show $[m^a]+i-[([m^a]+i-1)^\tau]>[(m-1)^a]+i.$

Noticing
\begin{align*}
[m^a]-[(m-1)^a] =am^{a-1}+o(m^{a-1})~~\mbox{as}~~m\rightarrow\infty
\end{align*}
and  $\tau\in (0,1-\frac1a),$ we find that as $m\rightarrow \infty$
\begin{align}\label{31}
\frac{[([m^a]+i-1)^{\tau}]}{[m^a]-[(m-1)^a]}
=O\left(\frac{m^{a\tau}}{m^{a-1}}\right)=O\left(m^{a\tau+1-a}\right)=o(1).
\end{align}
Thus, for fixed $i$ and $j$ the random vectors in the set
$\{\overline{\xi}_{k+1}:k+1\in I_j(i)\}$ are mutually independent
with possible exception of a finite number of vectors. Then by
noticing $\sup\limits_{k}E\|\overline{\xi}_k\|^2<\infty$ from
Proposition \ref{prop3} it follows that
\begin{align}
\sum\limits_{k=1}^{\infty}\frac{1}{k+1}\overline{\xi}_{k+1}<\infty~~\mathrm{a.s.}\label{32}
\end{align}

Further, we have
\begin{align}
\nonumber
&\left\|\sum\limits_{k=1}^{\infty}\frac{1}{k+1}(\xi_{k+1}-\overline{\xi}_{k+1})\right\|\\
\nonumber=&
\left\|\sum\limits_{k=1}^{\infty}\frac{1}{k+1}\sum\limits_{l=1}^{k-[k^{\tau}]}(1-\alpha_1)^{k+1-l}
(\Phi(k,l)\mathbf{1}-E\Phi(k,l)\mathbf{1})\right\|\\
=&O\left(\sum\limits_{k=1}^{\infty}\frac{1}{k+1}(1-\alpha_1)^{[k^{\tau}]}\right)=O(1),\label{33}
\end{align}
which combining with (\ref{32}) yields (\ref{29}). Thus, (\ref{24})
has been proved.

Noticing (\ref{18}), (\ref{23}), and (\ref{24}), by Proposition
\ref{prop1} we derive the assertion of Theorem 1.
\hfill$\blacksquare$

{\em Proof of Theorem 2:}  The proof can similarly be carried out as
that for Theorem 1 by Propositions \ref{prop2} and \ref{prop3}. We
only outline the key points.

First we have the exponential rate of convergence
$\|Ex_{1,k}-x^*\|=O(\rho^k)$ for some $0<\rho<1$ (for the analysis
we refer to, e.g., \cite{Tempo2010}).

By Proposition \ref{prop3} and carrying out a similar discussion as
that for (\ref{24}), (\ref{28}), and (\ref{29}), we can also prove
that
\begin{align}
\sum\limits_{k=0}^{\infty}
\frac{1}{(k+1)^s}(x_{1,k+1}-Ex_{1,k+1})<\infty~~\mathrm{a.s.}\label{35}
\end{align}
with $s>\frac{3}{2}-\frac{1}{a}$ for any fixed $a>1$, which implies
$s>\frac{1}{2}$.

Then by Proposition \ref{prop2}, we obtain (\ref{34}).
\hfill$\blacksquare$

\section{Extension to Case Where Multiple Pages Updating Simultaneously}

The protocol in Section II is based on the assumption that only one
page updates its PageRank value each time. In this section, we
discuss the convergence of DRPA for the case where multiple pages
update their PageRank values simultaneously. Notice that the problem
formulation in this section is precisely the same as that given in
\cite{Tempo2010}.

%\subsection{Multiple pages update simultaneously.}

%It is more practical to consider the case where multiple pages can
%update simultaneously and link failures randomly occur. The scheme
%is mathematically modeled as follows \cite{Tempo2010}.

Assume that the sequences of Bernoulli random variables
$\{\eta_i(k)\}_{k\geq0},~i=1,\cdots,n$ are mutually independent and
each sequence is of iid random variables with probabilities
\begin{align}
&\mathbb{P}\{\eta_i(k)=1\}=\beta,\label{36}\\
&\mathbb{P}\{\eta_i(k)=0\}=1-\beta,\label{37}
\end{align}
where $\beta\in(0,1]$. If $\eta_i(k)=1$, then page $i$ updates at
time $k,$  sending its PageRank value to the pages that page $i$ has
outgoing links to and requiring PageRank values from those pages
which page $i$ has incoming links from. While if $\eta_i(k)=0$, no
communication is required by page $i$.

Set $\eta(k)\triangleq(\eta_1(k),\cdots,\eta_n(k))$. The vector
$\eta(k)$ reflects  updating pages at time $k$. The corresponding
link matrix is given by
\begin{align}
(A_{p_1,\cdots,p_n})_{ij}\triangleq
\begin{cases}
a_{ij},~~\mathrm{if}~p_i=1~\mathrm{or}~p_j=1,\\
1-\sum\limits_{h:p_h=1}a_{hj},~~\mathrm{if}~p_i=0~\mathrm{and}~i=j,\\
0,~~\mathrm{if}~p_i=p_j=0~\mathrm{and}~i\neq j,
\end{cases}\label{40}
\end{align}
where $(p_1,\cdots,p_n)$ is a realization of $\eta(k)$. It is clear
that $A_{p_1,\cdots,p_n}$ is a sparse matrix.

Similar to (\ref{8}) and (\ref{12}), the DRPA for the multiple pages
updating is given by
\begin{align}
&x_{2,k+1}=(1-\alpha_2)A_{\eta(k)}x_{2,k}
+\frac{\alpha_2}{n}\mathbf{1},\label{38}\\
&\overline{x}_{2,k+1}=\frac{1}{k+1}\sum\limits_{l=0}^k
x_{2,l},\label{39}
\end{align}
where $x_{2,0}$ is an arbitrary probability vector and
$\alpha_2=\frac{\alpha(1-(1-\beta)^2)}{1-\alpha(1-\beta)^2}$.
Clearly, there are $2^n$ different link matrices.

\begin{remark}\label{rem4}
Define $M_2=(1-\alpha_2)EA_{\eta(k)}+\alpha_2\frac{S}{n}$. The
parameter
$\alpha_2=\frac{\alpha[1-(1-\beta)^2]}{1-\alpha(1-\beta)^2}$ is to
make $M_2$ to have the form: $M_2=\frac{\alpha_2}{\alpha}M+
\left(1-\frac{\alpha_2}{\alpha}\right)I$ (\cite{Tempo2010}). Thus
the eigenvector of $M_2$ corresponding to its biggest eigenvalue
equals the PageRank value $x^*$ which satisfies $x^*=Mx^*$. Noticing
that$\{A_{\eta(k)}\}$ is iid and carrying out the same discussion as
that for Theorems 1 and 2, we can show that
$\|\overline{x}_{2,k}-x^*\|
=o\left(\frac{1}{k^{\frac{1}{2}-\epsilon}}\right)~\mathrm{a.s.}$ for
any $\epsilon\in\left(0,\frac{1}{2}\right)$.
\end{remark}

\begin{remark}\label{rem5}
By carrying out the discussions similar to those done above, the
a.s. convergence of DRPA can be established for some other setups,
for example, when link failures randomly occur with the assumption
that the failure occurring is iid ([9]).
\end{remark}

\section{Concluding Remarks}

In the note, we have shown some further properties of DRPA
introduced in \cite{Tempo2010}. By using some convergence results of
SA, the strong consistency of estimates for the PageRank values as
well as their convergence rate are established for the cases where
one page updates each time and multiple pages update simultaneously.

There are many interesting problems left for future research. For
example, a key assumption for strong consistency of DRPA is that the
randomization law is iid. How to relax this assumption to the
dependent case, for example, to Markov chains, is interesting.
Another problem is to investigate the relation between the
convergence rate and the size $n$ of the web. It is also of interest
to consider the PageRank computation with communication delay and
web aggregation.

\end{document}